%% file: O1_long.tex
\documentclass[prd,twocolumn]{revtex4-1}
\usepackage[]{graphicx} 
\usepackage{epstopdf} 
\usepackage[]{units} 
\usepackage[]{color} 
\usepackage[]{amssymb} 
\usepackage[]{amsmath} 
\usepackage[]{hyperref} 
\usepackage[]{breakurl}
\usepackage[caption=false]{subfig} 
\usepackage[UKenglish]{babel}
\usepackage{multirow}
\usepackage{lineno}

\newcommand{\OOneStart}{{\color{black}September 12, 2015}}
\newcommand{\OOneStop}{{\color{black}January 19, 2016}}
\newcommand{\OOneStartShort}{{\color{black}September 2015}}
\newcommand{\OOneStopShort}{{\color{black}January 2016}}
\newcommand{\OOneLivetime}{{\color{black}49 days}}

\newcommand{\OOneCoincTime}{{\color{black}40\%}}

\newcommand{\msuncd}{$\mathrm{M_{\odot} c^2}$}
\newcommand{\msun}{$\mathrm{M_{\odot}}$}
\begin{document}

\title{All-sky search for long-duration gravitational wave transients in the first Advanced LIGO observing run}



\input{LSC_Aug2016_Virgo_Aug2016-prd.tex}


\date{\today}

\begin{abstract}
We present the results of a search for long-duration gravitational wave transients
in the data of the LIGO Hanford and LIGO Livingston second generation detectors between
\OOneStartShort~ and \OOneStopShort, with a total observational time of \OOneLivetime.
The search targets gravitational wave transients of \unit[10 -- 500]{s} duration in a
frequency band of \unit[24 -- 2048]{Hz}, with minimal assumptions about the signal waveform,
polarization, source direction, or time of occurrence. No significant events were observed.
As a result we set 90\% confidence upper limits on the rate
of long-duration gravitational wave transients for different types of gravitational wave
signals. We also show that the search is sensitive to sources in the Galaxy emitting at least
$\sim$ \unit[$10^{-8}$]{\msuncd} in gravitational waves.

\end{abstract}


\maketitle


\input{intro}

\input{dataset}

\input{searches}
\input{sensitivity}

\input{discussion}
\input{conclusion}

\begin{acknowledgments}
\input{LVCacknowledgments.tex}
\end{acknowledgments}

\bibliography{biblio}


\end{document}

%% file: LSC_Aug2016_Virgo_Aug2016-prd.tex



\author{%
B.~P.~Abbott,$^{1}$  
R.~Abbott,$^{1}$  
T.~D.~Abbott,$^{2}$  
M.~R.~Abernathy,$^{3}$  
F.~Acernese,$^{4,5}$ 
K.~Ackley,$^{6}$  
C.~Adams,$^{7}$  
T.~Adams,$^{8}$ 
P.~Addesso,$^{9}$  
R.~X.~Adhikari,$^{1}$
% ...
\\
\medskip
(LIGO Scientific Collaboration and Virgo Collaboration) 
\\
\medskip
{{}$^{\dag}$Deceased, March 2016. }%
}\email{lvc.publications@ligo.org}\noaffiliation
\affiliation {LIGO, California Institute of Technology, Pasadena, CA 91125, USA }
\affiliation {Louisiana State University, Baton Rouge, LA 70803, USA }
\affiliation {American University, Washington, D.C. 20016, USA }
\affiliation {Universit\`a di Salerno, Fisciano, I-84084 Salerno, Italy }
\affiliation {INFN, Sezione di Napoli, Complesso Universitario di Monte S.Angelo, I-80126 Napoli, Italy }
\affiliation {University of Florida, Gainesville, FL 32611, USA }
\affiliation {LIGO Livingston Observatory, Livingston, LA 70754, USA }
\affiliation {Laboratoire d'Annecy-le-Vieux de Physique des Particules (LAPP), Universit\'e Savoie Mont Blanc, CNRS/IN2P3, F-74941 Annecy-le-Vieux, France }
\affiliation {University of Sannio at Benevento, I-82100 Benevento, Italy and INFN, Sezione di Napoli, I-80100 Napoli, Italy }
\affiliation {Albert-Einstein-Institut, Max-Planck-Institut f\"ur Gravi\-ta\-tions\-physik, D-30167 Hannover, Germany }
\affiliation {Nikhef, Science Park, 1098 XG Amsterdam, The Netherlands }
\affiliation {LIGO, Massachusetts Institute of Technology, Cambridge, MA 02139, USA }
\affiliation {Instituto Nacional de Pesquisas Espaciais, 12227-010 S\~{a}o Jos\'{e} dos Campos, S\~{a}o Paulo, Brazil }
\affiliation {INFN, Gran Sasso Science Institute, I-67100 L'Aquila, Italy }
\affiliation {INFN, Sezione di Roma Tor Vergata, I-00133 Roma, Italy }
\affiliation {Inter-University Centre for Astronomy and Astrophysics, Pune 411007, India }
\affiliation {International Centre for Theoretical Sciences, Tata Institute of Fundamental Research, Bengaluru 560089, India }
\affiliation {University of Wisconsin-Milwaukee, Milwaukee, WI 53201, USA }
\affiliation {Leibniz Universit\"at Hannover, D-30167 Hannover, Germany }
\affiliation {Universit\`a di Pisa, I-56127 Pisa, Italy }
\affiliation {INFN, Sezione di Pisa, I-56127 Pisa, Italy }
\affiliation {Australian National University, Canberra, Australian Capital Territory 0200, Australia }
\affiliation {California State University Fullerton, Fullerton, CA 92831, USA }
\affiliation {LAL, Univ. Paris-Sud, CNRS/IN2P3, Universit\'e Paris-Saclay, F-91898 Orsay, France }
\affiliation {Chennai Mathematical Institute, Chennai 603103, India }
\affiliation {Universit\`a di Roma Tor Vergata, I-00133 Roma, Italy }
\affiliation {Universit\"at Hamburg, D-22761 Hamburg, Germany }
\affiliation {INFN, Sezione di Roma, I-00185 Roma, Italy }
\affiliation {Albert-Einstein-Institut, Max-Planck-Institut f\"ur Gravitations\-physik, D-14476 Potsdam-Golm, Germany }
\affiliation {APC, AstroParticule et Cosmologie, Universit\'e Paris Diderot, CNRS/IN2P3, CEA/Irfu, Observatoire de Paris, Sorbonne Paris Cit\'e, F-75205 Paris Cedex 13, France }
\affiliation {West Virginia University, Morgantown, WV 26506, USA }
\affiliation {Center for Gravitational Waves and Cosmology, West Virginia University, Morgantown, WV 26505, USA }
\affiliation {Universit\`a di Perugia, I-06123 Perugia, Italy }
\affiliation {INFN, Sezione di Perugia, I-06123 Perugia, Italy }
\affiliation {European Gravitational Observatory (EGO), I-56021 Cascina, Pisa, Italy }
\affiliation {Syracuse University, Syracuse, NY 13244, USA }
\affiliation {SUPA, University of Glasgow, Glasgow G12 8QQ, United Kingdom }
\affiliation {LIGO Hanford Observatory, Richland, WA 99352, USA }
\affiliation {Wigner RCP, RMKI, H-1121 Budapest, Konkoly Thege Mikl\'os \'ut 29-33, Hungary }
\affiliation {Columbia University, New York, NY 10027, USA }
\affiliation {Stanford University, Stanford, CA 94305, USA }
\affiliation {Universit\`a di Padova, Dipartimento di Fisica e Astronomia, I-35131 Padova, Italy }
\affiliation {INFN, Sezione di Padova, I-35131 Padova, Italy }
\affiliation {Nicolaus Copernicus Astronomical Center, Polish Academy of Sciences, 00-716, Warsaw, Poland }
\affiliation {Center for Relativistic Astrophysics and School of Physics, Georgia Institute of Technology, Atlanta, GA 30332, USA }
\affiliation {University of Birmingham, Birmingham B15 2TT, United Kingdom }
\affiliation {Universit\`a degli Studi di Genova, I-16146 Genova, Italy }
\affiliation {INFN, Sezione di Genova, I-16146 Genova, Italy }
\affiliation {RRCAT, Indore MP 452013, India }
\affiliation {Faculty of Physics, Lomonosov Moscow State University, Moscow 119991, Russia }
\affiliation {SUPA, University of the West of Scotland, Paisley PA1 2BE, United Kingdom }
\affiliation {Caltech CaRT, Pasadena, CA 91125, USA }
\affiliation {University of Western Australia, Crawley, Western Australia 6009, Australia }
\affiliation {Department of Astrophysics/IMAPP, Radboud University Nijmegen, P.O. Box 9010, 6500 GL Nijmegen, The Netherlands }
\affiliation {Artemis, Universit\'e C\^ote d'Azur, CNRS, Observatoire C\^ote d'Azur, CS 34229, F-06304 Nice Cedex 4, France }
\affiliation {Institut de Physique de Rennes, CNRS, Universit\'e de Rennes 1, F-35042 Rennes, France }
\affiliation {Washington State University, Pullman, WA 99164, USA }
\affiliation {Universit\`a degli Studi di Urbino 'Carlo Bo', I-61029 Urbino, Italy }
\affiliation {INFN, Sezione di Firenze, I-50019 Sesto Fiorentino, Firenze, Italy }
\affiliation {University of Oregon, Eugene, OR 97403, USA }
\affiliation {Laboratoire Kastler Brossel, UPMC-Sorbonne Universit\'es, CNRS, ENS-PSL Research University, Coll\`ege de France, F-75005 Paris, France }
\affiliation {Carleton College, Northfield, MN 55057, USA }
\affiliation {Astronomical Observatory Warsaw University, 00-478 Warsaw, Poland }
\affiliation {VU University Amsterdam, 1081 HV Amsterdam, The Netherlands }
\affiliation {University of Maryland, College Park, MD 20742, USA }
\affiliation {Laboratoire des Mat\'eriaux Avanc\'es (LMA), CNRS/IN2P3, F-69622 Villeurbanne, France }
\affiliation {Universit\'e Claude Bernard Lyon 1, F-69622 Villeurbanne, France }
\affiliation {Universit\`a di Napoli 'Federico II', Complesso Universitario di Monte S.Angelo, I-80126 Napoli, Italy }
\affiliation {NASA/Goddard Space Flight Center, Greenbelt, MD 20771, USA }
\affiliation {RESCEU, University of Tokyo, Tokyo, 113-0033, Japan. }
\affiliation {University of Adelaide, Adelaide, South Australia 5005, Australia }
\affiliation {Tsinghua University, Beijing 100084, China }
\affiliation {Texas Tech University, Lubbock, TX 79409, USA }
\affiliation {The University of Mississippi, University, MS 38677, USA }
\affiliation {The Pennsylvania State University, University Park, PA 16802, USA }
\affiliation {National Tsing Hua University, Hsinchu City, 30013 Taiwan, Republic of China }
\affiliation {Charles Sturt University, Wagga Wagga, New South Wales 2678, Australia }
\affiliation {University of Chicago, Chicago, IL 60637, USA }
\affiliation {Kenyon College, Gambier, OH 43022, USA }
\affiliation {Korea Institute of Science and Technology Information, Daejeon 305-806, Korea }
\affiliation {University of Cambridge, Cambridge CB2 1TN, United Kingdom }
\affiliation {Universit\`a di Roma 'La Sapienza', I-00185 Roma, Italy }
\affiliation {Universit\'e Libre de Bruxelles, Brussels 1050, Belgium }
\affiliation {Sonoma State University, Rohnert Park, CA 94928, USA }
\affiliation {Montana State University, Bozeman, MT 59717, USA }
\affiliation {Center for Interdisciplinary Exploration \& Research in Astrophysics (CIERA), Northwestern University, Evanston, IL 60208, USA }
\affiliation {Universitat de les Illes Balears, IAC3---IEEC, E-07122 Palma de Mallorca, Spain }
\affiliation {The University of Texas Rio Grande Valley, Brownsville, TX 78520, USA }
\affiliation {Bellevue College, Bellevue, WA 98007, USA }
\affiliation {Institute for Plasma Research, Bhat, Gandhinagar 382428, India }
\affiliation {The University of Sheffield, Sheffield S10 2TN, United Kingdom }
\affiliation {California State University, Los Angeles, 5154 State University Dr, Los Angeles, CA 90032, USA }
\affiliation {Universit\`a di Trento, Dipartimento di Fisica, I-38123 Povo, Trento, Italy }
\affiliation {INFN, Trento Institute for Fundamental Physics and Applications, I-38123 Povo, Trento, Italy }
\affiliation {Cardiff University, Cardiff CF24 3AA, United Kingdom }
\affiliation {Montclair State University, Montclair, NJ 07043, USA }
\affiliation {National Astronomical Observatory of Japan, 2-21-1 Osawa, Mitaka, Tokyo 181-8588, Japan }
\affiliation {Canadian Institute for Theoretical Astrophysics, University of Toronto, Toronto, Ontario M5S 3H8, Canada }
\affiliation {MTA E\"otv\"os University, ``Lendulet'' Astrophysics Research Group, Budapest 1117, Hungary }
\affiliation {School of Mathematics, University of Edinburgh, Edinburgh EH9 3FD, United Kingdom }
\affiliation {University and Institute of Advanced Research, Gandhinagar, Gujarat 382007, India }
\affiliation {IISER-TVM, CET Campus, Trivandrum Kerala 695016, India }
\affiliation {University of Szeged, D\'om t\'er 9, Szeged 6720, Hungary }
\affiliation {Embry-Riddle Aeronautical University, Prescott, AZ 86301, USA }
\affiliation {Tata Institute of Fundamental Research, Mumbai 400005, India }
\affiliation {INAF, Osservatorio Astronomico di Capodimonte, I-80131, Napoli, Italy }
\affiliation {University of Michigan, Ann Arbor, MI 48109, USA }
\affiliation {Rochester Institute of Technology, Rochester, NY 14623, USA }
\affiliation {NCSA, University of Illinois at Urbana-Champaign, Urbana, IL 61801, USA }
\affiliation {University of Bia{\l }ystok, 15-424 Bia{\l }ystok, Poland }
\affiliation {SUPA, University of Strathclyde, Glasgow G1 1XQ, United Kingdom }
\affiliation {University of Southampton, Southampton SO17 1BJ, United Kingdom }
\affiliation {University of Washington Bothell, 18115 Campus Way NE, Bothell, WA 98011, USA }
\affiliation {Institute of Applied Physics, Nizhny Novgorod, 603950, Russia }
\affiliation {Seoul National University, Seoul 151-742, Korea }
\affiliation {Inje University Gimhae, 621-749 South Gyeongsang, Korea }
\affiliation {National Institute for Mathematical Sciences, Daejeon 305-390, Korea }
\affiliation {Pusan National University, Busan 609-735, Korea }
\affiliation {NCBJ, 05-400 \'Swierk-Otwock, Poland }
\affiliation {Institute of Mathematics, Polish Academy of Sciences, 00656 Warsaw, Poland }
\affiliation {The School of Physics \& Astronomy, Monash University, Clayton 3800, Victoria, Australia }
\affiliation {Hanyang University, Seoul 133-791, Korea }
\affiliation {The Chinese University of Hong Kong, Shatin, NT, Hong Kong }
\affiliation {University of Alabama in Huntsville, Huntsville, AL 35899, USA }
\affiliation {ESPCI, CNRS, F-75005 Paris, France }
\affiliation {University of Minnesota, Minneapolis, MN 55455, USA }
\affiliation {Universit\`a di Camerino, Dipartimento di Fisica, I-62032 Camerino, Italy }
\affiliation {Southern University and A\&M College, Baton Rouge, LA 70813, USA }
\affiliation {The University of Melbourne, Parkville, Victoria 3010, Australia }
\affiliation {College of William and Mary, Williamsburg, VA 23187, USA }
\affiliation {Instituto de F\'\i sica Te\'orica, University Estadual Paulista/ICTP South American Institute for Fundamental Research, S\~ao Paulo SP 01140-070, Brazil }
\affiliation {Whitman College, 345 Boyer Avenue, Walla Walla, WA 99362 USA }
\affiliation {Universit\'e de Lyon, F-69361 Lyon, France }
\affiliation {Hobart and William Smith Colleges, Geneva, NY 14456, USA }
\affiliation {Janusz Gil Institute of Astronomy, University of Zielona G\'ora, 65-265 Zielona G\'ora, Poland }
\affiliation {King's College London, University of London, London WC2R 2LS, United Kingdom }
\affiliation {IISER-Kolkata, Mohanpur, West Bengal 741252, India }
\affiliation {Indian Institute of Technology, Gandhinagar Ahmedabad Gujarat 382424, India }
\affiliation {Andrews University, Berrien Springs, MI 49104, USA }
\affiliation {Universit\`a di Siena, I-53100 Siena, Italy }
\affiliation {Trinity University, San Antonio, TX 78212, USA }
\affiliation {University of Washington, Seattle, WA 98195, USA }
\affiliation {Abilene Christian University, Abilene, TX 79699, USA }


%% file: intro.tex
\section{Introduction}

The first observing runs of the Advanced LIGO and Advanced Virgo detectors, with significant sensitivity improvements compared to the first generation detectors, yielded in less than two years incredible discoveries and major astrophysics results via gravitational wave (GW) detections. The first observed GW signals corresponded to the final moments of the coalescence of two stellar-mass black holes and their final plunge. GW150914 and GW151226 were observed with high confidence ($> 5\sigma $), while LVT151012 was identified with a lower significance ($1.7 \sigma$)~\cite{Abbott:2016blz,AbEA2016b,Abbott:2016pea,Abbott:2016uux,TheLIGOScientific:2016qqj} during the first observing run (O1). During the second observing run (O2), GW170104 and GW170814 (which was detected simultaneously by the three LIGO and Virgo detectors) have confirmed the estimated rate of stellar-mass black hole mergers~\cite{Abbott:2017vtc,Abbott:2017oio}. Lastly, the observation of a binary neutron star inspiral by the LIGO and Virgo network~\cite{Abbott:2017qsa} in association with a gamma-ray burst~\cite{Abbott:2017mdv} and a multitude of broadband electromagnetic counterpart observations~\cite{GBM:2017lvd} has opened up a new era in multimessenger astronomy.

The searches that observed these binary compact object systems were also targetting neutron star -- black hole mergers~\cite{Abbott:2016ymx,Abbott:2016ezn} as well as intermediate-mass black hole mergers of total mass up to 600~\msun~\cite{Abbott:2017iws}. So far, only O1 observing run results have been reported for these sources, and no other compact binary coalescence, nor any short duration signal targeted by unmodeled short duration searches~\cite{Abbott:2016ezn} have been observed.


In this paper, we present an all-sky search for unmodeled long-duration (\unit[10--500]{s}) transient GW events. Astrophysical compact sources undergoing complex dynamics and hydrodynamic instabilities are expected to emit long-lasting GWs. For example, fallback accretion onto a newborn neutron star can lead to a non-axisymmetric deformation which emits GWs until the neutron star collapses to a black hole~\cite{Lai:1994ke,Cutler:2002nw,piro:11,PiTh2012}. Non-axisymmetric accretion disc fragmentation and instabilities can lead to material spiraling into the central stellar-mass black hole, emitting GWs~\cite{Piro:2006ja,VaPu2001,VaPu2008}. Long-duration GWs may also be emitted by non-axisymmetric deformations in magnetars~\cite{corsi:09,gualteri:11}, which are possible progenitors of long and short GRBs~\cite{Metzger:2010pp,rowlinson:13}. Finally, core-collapse supernovae simulations have shown that the turbulent and chaotic fluid movements that occur in the proto-neutron star formed a few hundred milliseconds after the core collapse can excite long-lasting surface g-modes whose frequency drifts over time~\cite{Marek:2008qi,Murphy:2009dx}.

We extend the search for long-duration GW transients previously carried out on initial LIGO data from the period 2005--2010~\cite{Abbott:2015vir}. Four pipelines have been used to double the frequency band coverage from (\unit[40--1000]{Hz}) to (\unit[24--2048]{Hz}), and new waveform models have been used to estimate the pipelines' sensitivities. We explicitly demonstrate that the search is capable of efficiently detecting three of the four potential sources mentioned above. No significant events were detected and consequently, upper limits have been set on the rate of long-duration transient signals.
 
The organization of the paper is as follows. In Section~\ref{sec:dataset}, we describe the dataset. Section~\ref{sec:searches} is devoted to a brief description of the pipelines, whose sensitivities are presented in Section~\ref{sec:sensitivity}. In Section~\ref{sec:discussion}, we give and discuss the results, then we conclude in Section~\ref{sec:conclusion} with a discussion of future expectations.

%% file: dataset.tex
\section{Data set \& data quality}\label{sec:dataset}

This O1 analysis uses data from \OOneStart~to \OOneStop. The LIGO detectors in Hanford, WA and Livingston, LA ran with \OOneCoincTime~coincident time. For this long-duration transient search, about two days of coincident data have been discarded because they were affected by major detector failures or problematic weather conditions. The remaining \OOneLivetime~ of coincident data still contain many non-stationary short duration noise events that can mimic a signal. These noise events, or ``glitches'', have a multitude of causes. For instance, low frequency glitches are caused by surges in power lines or seismic events, while many high frequency glitches are caused by resonances in the test mass suspension wires~\cite{Abbott:2016zmo}. Many of these effects can be tracked in auxiliary sensors that we use to define the severity of the loss of data quality~\cite{Christensen:2010zz,Aasi:2014mqd,Abbott:2016zmo}.

The signals targeted by the long-duration transients search have their energy spread over a large time span. Consequently, even modest excesses of noise directly influence the signal reconstruction. In order to be considered as a potential real signal, events must be seen coincidently in the two LIGO detectors. This requirement eliminates most of the noise events due to glitches. An accurate background estimation using the data themselves is therefore necessary to measure the significance of any coincident excess of energy. A false alarm rate (FAR) is estimated after safe veto methods are applied to get rid of as many glitch events as possible. While a few families of these noise events can be suppressed by vetoes based on auxiliary channels, each search pipeline has its own background reduction strategy and its own implementation of the time-slides method~\cite{0264-9381-27-1-015005} to estimate the FAR. It consists in introducing a time-shift in one detector's strain time series. Details on these topics are provided in the next section.

%% file: searches.tex
\section{Pipelines}\label{sec:searches}
Four pipelines are used to analyze the data set and search for long-duration GW transient signals. These pipelines are described in the sub-sections that follow.

\subsection{Coherent WaveBurst}
Coherent WaveBurst (cWB) is a pipeline designed to search for generic GW transients.
Using a maximum-likelihood-ratio statistic~\cite{Klimenko:2015ypf}, it identifies coincident excess power events (triggers) in a time-frequency space.
The long-duration transient cWB search is implemented with the same pipeline also used to search for short transient events~\cite{Abbott:2016ezn} with a few specific changes:
It operates in the frequency range \unit[24--2048]{Hz} and only data which pass the strictest 
data quality criteria are examined (see Section \ref{sec:dataset} and~\cite{Abbott:2016ezn}). Events are ranked according to their detection statistic ($\eta_{c}$), which is related to the event signal-to-noise ratio ($\mathrm{SNR}$). A primary selection is based on the network correlation coefficient $C_{c}$~\cite{Klimenko:2015ypf}, which measures the degree of correlation between the detectors, and the energy-weighted duration of detected triggers.
Events with $C_{c}<0.6$ or duration \unit[$<1.5$]{s} are excluded from the analysis. The selection criterion based on duration is specific to this long-duration search and it is the most powerful selection criteria to suppress background triggers.
To characterize the FAR, the data of one interferometer is shifted in time (the so called time-slides method) with respect to the other interferometer by multiple delays of \unit[1]{s} for an equivalent total time of $\sim$ 70 years of coincident time.

\subsection{The STAMP-AS pipeline}
The all-sky STAMP-AS pipeline based on the Stochastic Transient Analysis Multi-detector Pipeline~\cite{Abbott:2015vir} cross-correlates data from two detectors and builds coherent time-frequency maps ($tf$-maps) of SNR with a pixel size of 1s $\times$ 1Hz. The SNR is computed for each second of data by estimating the mean noise over the neighboring seconds on each side. Pixels in frequency bins corresponding to known instrumental lines are suppressed. Once the $tf$-maps are built, overlapping clusters that pass a SNR threshold of 0.75 are grouped to form triggers. There are two variants of STAMP-AS that differ in cluster grouping strategy: Zebragard and Lonetrack.

\subsubsection{Zebragard}
Working with $tf$-maps of size \unit[24-2000]{Hz} $\times$ \unit[500]{s}, Zebragard groups together pixels above a given SNR threshold that lie within a 4 pixel distance from each other. Because a sub-optimal number of sky positions are targeted, a signal can be anti-coherent (negative SNR). The algorithm addresses this in such a way that the loss of efficiency due to the limited number of tested sky positions is less than 10\%~\cite{prestegard:2016}.
The trigger ranking statistic, $\Theta_{\Gamma}$, is defined as the quadratic sum of the SNR of the individual pixels. This analysis uses the same configuration and the same background rejection strategy against short-duration noise transient ``glitches'' (the fraction of SNR in each time bin must be smaller than 0.5) as in~\cite{Abbott:2015vir}. In addition, the O1 data set contains an excess of background triggers that required developing additional vetoes. For example, using the fact that the two LIGO detectors are almost aligned, triggers due to a loud glitch in one detector are suppressed by demanding that the SNR ratio between the two detectors is smaller than 3.
Mechanical resonances excited when the optical cavities of the interferometer arms are locked generate an excess of triggers at \unit[39]{Hz} and \unit[43]{Hz} at well identified times. Finally, the remaining glitches are efficiently suppressed by data quality vetoes based on auxiliary channels~\cite{1742-6596-243-1-012005}. It has been verified that these vetoes minimally affect the search for the targeted signals (less than 5\% of simulated signals are lost). The background is estimated by time-shifting the data of one detector relative to the other in steps of \unit[250]{s}. Data quality investigations and veto tuning are performed using a subset of the time-shifted triggers. The background rate is estimated with 600 time shift values between the detectors for an equivalent total time of $\sim$ 78 years of coincident time for the O1 data set. 

\subsubsection{Lonetrack}
Lonetrack uses seedless clustering to integrate the signal power along spectrogram tracks using templates chosen to capture the salient features of a wide class of signal models. Templates here are not meant to exactly match the signal but rather to identify a few isolated pixels that are part of the signals.
B\'ezier curves~\cite{Far1996,ThCo2013,ThCo2014,CoMe2015b,ThCo2015}, a post-Newtonian expansion for time-frequency track of circular compact binary coalescence signals~\cite{CoTh2014}, and an analytic expression for low-to-moderate eccentric compact binaries~\cite{CoMe2015} have all been used previously as seedless clustering parametrizations.
These parameterizations are used to create template banks of frequency-time tracks.
In this present search, B\'ezier curves were used in order to be sensitive to as many signal models as possible.

The Lonetrack search hierarchically selects the most promising triggers. This allows us to estimate the events' significance at very low FAR (to reach the equivalent of 5$\sigma$ detection probability). It begins by applying seedless clustering to analyze spectrograms of a single-detector, incoherent statistic~\cite{ThCo2015}.
For times that pass a threshold on SNR of 6, $tf$-maps of cross-power SNR are constructed using the tracks derived from the single detector, incoherent statistic. 
This analysis is carried out for 400 evenly spaced values of \unit[0.05]{ms} time delay between the detectors. The FAR is estimated with an equivalent total time of $\sim$ 12,000 years.
The detection statistic to rank triggers is the maximum SNR found per map.

\subsection{X-SphRad}
The X-pipeline Spherical Radiometer (X-SphRad) is a fast cross-correlator in the spherical harmonic domain~\cite{Cannon:2007br}. The spherical radiometry approach takes advantage of the fact that sky maps in GW searches show strong correlations over large angular scales in a pattern determined by the network geometry~\cite{edwards:2013}. Computing sky maps indirectly through their spherical harmonics minimizes the number of redundant calculations, allowing the data to be processed independently of sky position. The pipeline is built on X-pipeline~\cite{Sutton:2009gi,Was:2012zq} which whitens the data in the pre-processing step and then post-processes the event triggers output using the spherical radiometer. The pipeline uses the ratio of the power in the homogeneous polynomials of degree $l > 0$ modes to that in the $l=0$ mode to rank triggers. This ranking statistic provides a discriminatory power for rejecting background glitches~\cite{Edwards:2012zzc}.
To estimate the background, X-SphRad time-shifts the data for each detector in the network. The X-SphRad O1 search used an equivalent total time of $57$ years and covers the frequency band \unit[24--1000]{Hz}.

%% file: sensitivity.tex
\section{Sensitivity} \label{sec:sensitivity}
The sensitivity of each pipeline is estimated using 22 different types of simulated GW signals. Half of these are based on astrophysical source models and can be divided into 3 families: fallback accretion onto neutron stars (FA), black hole accretion disk instabilities (ADI) and magnetars. The other waveforms have ad-hoc morphologies that encapsulate the main characteristics for long-duration transients. The next section briefly describes the models of sources whose chosen parameters are given in table~\ref{table:LT_cmp}.

\subsection{Waveform descriptions}\label{sec:waveforms}
\textbf{FA}: The fallback accretion disk model~\cite{PiTh2012} focuses on newly born spinning neutron stars. In some unstable configurations, a non-axisymmetric deformation appears causing the production of GWs. The signal lasts from $\sim$ \unit[10]{s} up to a few \unit[100]{s} and its frequency evolution is almost linear.

\textbf{ADI}: This family includes five waveforms already considered and described in the LIGO S5/S6 search~\cite{Abbott:2015vir} and the O1 GRB search~\cite{Abbott:2016jsd}. In this model~\cite{VaPu2001,VaPu2008}, a thick accretion disk is coupled to a Kerr black hole through strong magnetic fields. This coupling is thought to generate turbulence in the accretion torus that may form clumps of matter. The quadrupole components of the disk lead to gravitational wave emission that spin down the black hole and separate the clumps. The anti-chirp like waveform (frequency and amplitude decreases over time) depends on the mass of the central black hole $M$, the Kerr spin parameter $a_{*}$, and the fraction $\epsilon$ of the disk mass that forms clumps. 

\textbf{Magnetar}: Magnetic deformation of a rapidly rotating neutron star can generate long-lasting GWs that can live up to one hour with a slowly decreasing frequency and amplitude (i.e., an anti-chirp). We used the model described in~\cite{DaGi2015}, which includes two parameters: the magnetic ellipticity $\epsilon_{b}$ and the spin frequency $f_{0}$ of the neutron star, that entirely describe the frequency and amplitude variations.

\textbf{Ad-hoc waveforms}: These include monochromatic waveforms (MONO) and waveforms with a linear (LINE) and quadratic (QUAD) frequency evolution, as well as white noise band-limited (WNB) and sine-Gaussian bursts (SG)~\cite{Abbott:2015vir}. All of these waveforms have duration from $\sim$ \unit[10]{s} up to a few \unit[100]{s} and frequencies spanning the analysis range.

\subsection{Detection efficiencies}
In order to determine the detection efficiencies, waveforms have been added coherently to the detector data at randomly chosen times over the full run period. We are using waveforms ($H_{+}$ and $H_{\times}$ polarizations) that have been generated in the frame of the source. For each chosen time we draw a source sky position such that the whole set of source positions is uniformly distributed over the sky. In the frame of the detector the waveforms are elliptically polarized with an ellipticity that varies uniformally between 0 and 1. The waveform amplitudes are also varied in order to estimate the dependency of the efficiency on the strength of the signal at a given FAR. Efficiency is simply the fraction of signals that are detected with a ranking statistic equal or larger than a value corresponding to the given FAR. To measure the intrinsic amplitude strength of a waveform, we use the root-sum-square strain amplitude at the Earth $h_{\mathrm{rss}}$ defined as in~\cite{Abbott:2015vir},
\begin{equation}
  h_{\mathrm{rss}}=\int_{-\infty}^{\infty} (H_{+}^2(t)+H_{\times}^2(t))~ dt,
\end{equation}
where $H_{+}$ and $H_{\times}$ are the GW strain polarizations in the source frame.
Table~\ref{table:LT_cmp} provides the values of $h_{\mathrm{rss}}$ at which each pipeline recovers 50\% or fewer of the injected signals for a FAR of 1 event in 50 years. 
Generally, cWB, Zebragard and X-SphRad have similar sensitivities while Lonetrack is better by a factor 2 for the waveforms that are well fit by B\'ezier curves (LINE and QUAD).

Some of the listed waveforms are not detectable by a given pipeline. This is naturally the case for $>$ \unit[1]{kHz} signals for X-SphRad. But this is also the case for monochromatic signals (MONO and SG) for cWB and Lonetrack. The reasons are different for each pipeline. For example, the way the pipelines whiten the data or estimate the detector noise power spectrum may wash out continuous signals. This is the case for cWB and to a lesser extent for Zebragard and X-SphRad. Lonetrack by construction has no sensitivity to monochromatic signals and band limited white noise as these types of waveforms are not modelled by a Bezier curve. 

\begin{figure*}[hbtp!]
  \centering
  \includegraphics[height=5in]{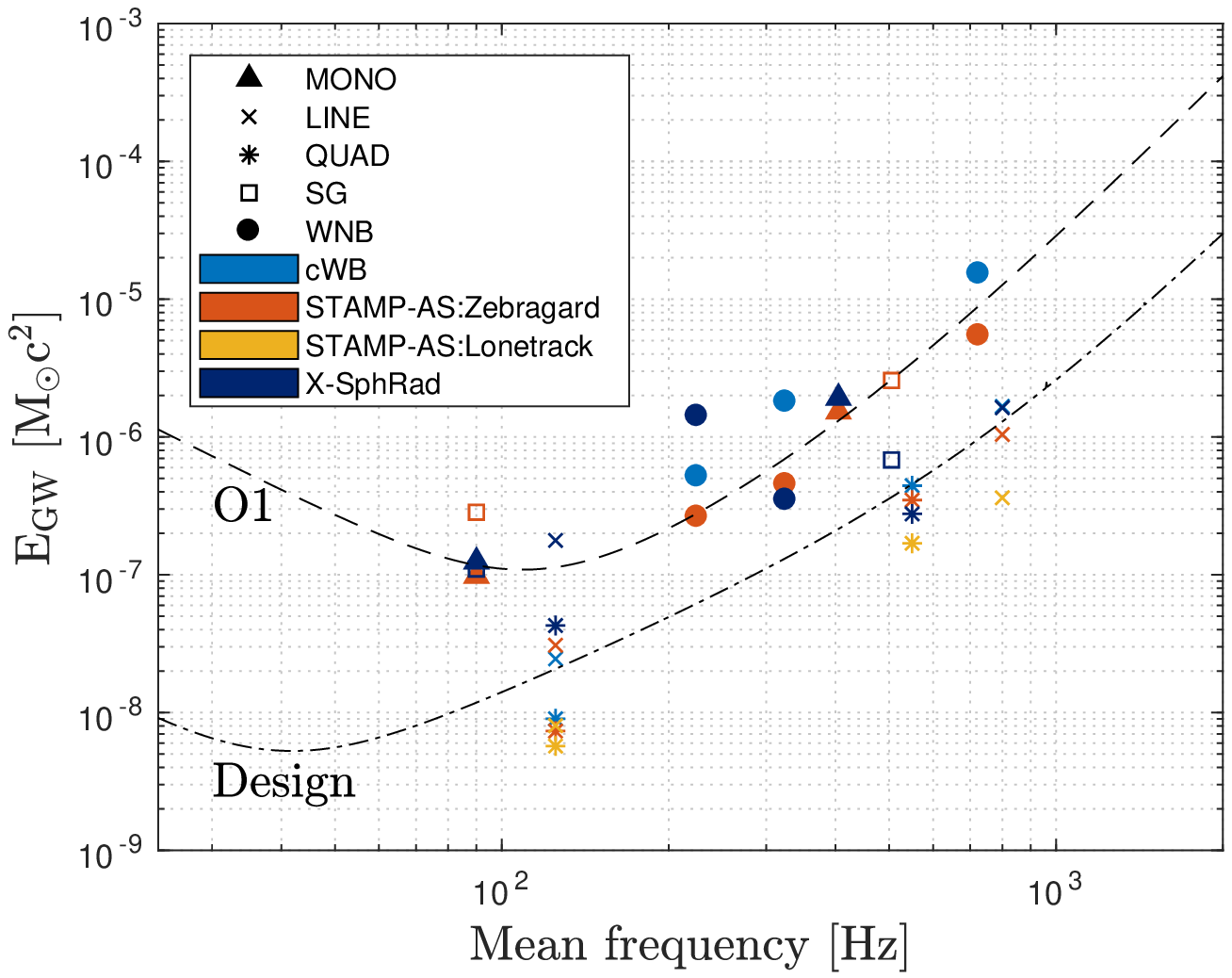}
  \caption{Emitted GW energy versus frequency for sources located at \unit[10]{kpc} detected with 50\% efficiency and a FAR of 1 event in 50 years. Results are shown for all the ad-hoc waveforms. The ``O1'' and ``Design'' curves are obtained with a monochromatic signal single template matched filtering search using the measured O1 and the predicted high-power signal recycling zero-detuning Advanced LIGO~\cite{aLIGO_sensitivity} sensitivity curves respectively. Both curves are rescaled so that the curve ``O1'' matches the MONO results of this search.}
  \label{fig:adhoc_energy}
\end{figure*}

Figure~\ref{fig:adhoc_energy} displays the GW energy emitted by a source located at \unit[10]{kpc} for which the search efficiency drops below 50\% for a FAR of 1 event in 50 years. The energy provides a universal quantity that can be directly compared to astrophysical predictions of the different possible sources. Assuming an isotropic GW emission, the energy emitted by a source at a distance $r$ is given by
\begin{equation}
E_{\mathrm{GW}} = \frac{c^{3}r^{2}}{4G} \int_{-\infty}^{\infty} \langle\dot{h}_{+}^{2}+\dot{h}_{\times}^{2}\rangle~ dt ,
\label{eq:energy}
\end{equation}
where $\dot{h}_{+}$ and $\dot{h}_{\times}$ are the time derivative of the GW strain for the two polarizations in the detector frame. 
For the sake of visibility, only ad-hoc model waveforms are considered in the figure while values for all waveforms are reported in Table~\ref{table:LT_cmp}. It illustrates the dependence on the signal frequency which roughly follows the detectors' sensitivity. Yet, one also sees that monochromatic (MONO and SG) and band limited white noise (WNB) waveform detections are systematically less efficient than the other waveforms. The minimal GW energy emitted by a source detected in the Galaxy (\unit[10]{kpc}) is of the order of a few \unit[$10^{-8}$]{\msuncd}. If one now looks at each pipeline's performance, for a given type of source, the detectable GW energy is spread over almost one order of magnitude and the most sensitive pipeline is different for each source. 

To project the search sensitivity forward to the Advanced LIGO detectors design sensitivity, we have considered the matched filtered search results for an idealized monochromatic signal with a detection SNR threshold of 8. We are using monochromatic signals because the frequency is well defined.
Results are then rescaled with a single factor such that the ``O1'' curve approximatively matches the MONO results of the O1 search. The ``Design'' curve is obtained using the predicted design Advanced LIGO high-power signal recycling zero-detuning sensitivity curves~\cite{aLIGO_sensitivity}, rescaled using the same factor as the ``O1'' curve. These curves show how the sensitivity to monochromatic signals will evolve through the future observing runs assuming a FAR of 1 event in 50 years. In particular, a gain of two orders of magnitude on the energy is expected at low frequency with the Advanced LIGO design sensitivity. Similar trends for the other waveforms are expected.

\input{tabsens}

%% file: tabsens.tex
\renewcommand{\arraystretch}{0.5}
\begin{table*}
\begin{tabular}{c|ccc|cccc|cccc}
  \hline
  \hline
\multirow{3}{*}{Waveform}    & \multicolumn{3}{c}{Properties} & \multicolumn{4}{c}{$h_{\mathrm{rss}}^{50\%}$ [1e-21 Hz$^{1/2}$] } &  \multicolumn{4}{c}{$E_{\mathrm{GW}}^{50\%}$ [\msuncd] } \\
\cline{2-12}                 & \footnotesize Parameters & \footnotesize Duration & \footnotesize Frequency & \footnotesize cWB &  \multicolumn{2}{c}{\footnotesize STAMP-AS} & \footnotesize X-SphRad    & \footnotesize cWB &  \multicolumn{2}{c}{\footnotesize STAMP-AS} &  \footnotesize X-SphRad \\
\cline{6-7}\cline{10-11}     &        & \footnotesize [s]  & \footnotesize [Hz]  &     & \footnotesize Zebragard     & \footnotesize Lonetrack    &             &     & \footnotesize Zebragard & \footnotesize Lonetrack &   \\ 
\hline
\hline
FA A      & -     &  25 & 1200-1500 &     2.55  & 2.05 &     1.62 & -        & 1.32e-05 & 8.49e-06 & 5.36e-06 & - \\
\hline
FA B      & -     & 197 &  800-1075 &     2.19  & 2.02 &     1.16 & -        & 4.77e-06 & 4.04e-06 & 1.34e-06 & - \\
\hline
\hline
ADI A     & $M$=\unit[5]{\msun}     &  39 &   135-166 &    0.48  & 0.54 &     0.42 & 0.39 & 5.84e-09 & 7.32e-09 & 4.43e-09 & 3.83e-09\\ 
          & $a_{*}=0.3$       &&&&&&&&&&\\
          & $\epsilon=0.05$  &&&&&&&&&&\\
\hline
ADI B     & $M$=\unit[10]{\msun}    &   9 &   110-209 &    0.51  & 0.55 &     0.57 & 0.52 & 6.45e-09 & 7.35e-09 & 7.98e-09 & 7.43e-09\\ 
          & $a_{*}=0.95$      &&&&&&&&&&\\
          & $\epsilon=0.2$   &&&&&&&&&&\\
\hline
ADI C     & $M$=\unit[10]{\msun}    & 236 &   130-251 &    1.07  & 1.02 &     0.76 & 1.38 & 2.97e-08 & 2.71e-08 & 1.49e-08 & 4.91e-08\\ 
          & $a_{*}=0.95$      &&&&&&&&&&\\
          & $\epsilon=0.04$  &&&&&&&&&&\\
\hline
ADI D     & $M$=\unit[3]{\msun}     & 142 &   119-173 &    0.86  & 1.04 &     0.70 & 1.08 & 1.66e-08 & 2.45e-08 & 1.12e-08 & 2.65e-08\\ 
          & $a_{*}=0.7$       &&&&&&&&&&\\
          & $\epsilon=0.035$  &&&&&&&&&&\\
\hline
ADI E     & $M$=\unit[8]{\msun}     &  76 &   111-260 &    0.75  & 0.64 &     0.55 & 1.31 & 1.51e-09 & 1.11e-09 & 8.10e-09 & 4.68e-08\\ 
          & $a_{*}=0.99$      &&&&&&&&&&\\
          & $\epsilon=0.065$  &&&&&&&&&&\\
\hline
\hline
magnetar D& $\epsilon_b=0.005$ & 400 & 1598-1900 &     5.07  & 6.72 &     3.70 & -        & 4.62e-05 & 8.12e-05 & 2.49e-05 & - \\
          & $f_{0}$=\unit[1598]{Hz}      &&&&&&&&&&\\
\hline
magnetar E& $\epsilon_b=0.01$  & 400 & 1171-1450 &     3.99  & 3.94 &     2.11 & -        & 2.14e-05 & 2.09e-05 & 5.97e-06 & - \\
          & $f_{0}$=\unit[1171]{Hz}      &&&&&&&&&&\\
\hline
magnetar F& $\epsilon_b=0.5$   & 400 &   579-950 &     2.46  & 2.09 &     1.18 & 1.75 & 3.40e-06 & 2.46e-06 & 7.79e-07 & 1.73e-06\\
          & $f_{0}$=\unit[579]{Hz}       &&&&&&&&&&\\
\hline
magnetar G& $\epsilon_b=0.08$  & 400 &   400-490 &     1.72  & 2.14 &     1.22 & 1.04 & 6.40e-07 & 9.89e-07 & 3.18e-07 & 2.36e-07\\
          & $f_{0}$=\unit[405]{Hz}       &&&&&&&&&&\\
\hline
\hline
MONO A    & $f_{0}$=\unit[90]{Hz}             & 150 &  90 & -          & 3.28 & -        & 3.70 & -        & 9.80e-08 & -        & 1.24e-07\\ 
          & $\frac{df}{dt}$=0     &&&&&&&&&&\\  
          & $\frac{d^2f}{dt^2}$=0 &&&&&&&&&&\\
\hline
MONO C    & $f_{0}$=\unit[405]{Hz}            & 250 & 405 & -          & 2.92 & -        & 3.28 & -        & 1.52e-06 & -        & 1.92e-06  \\ 
          & $\frac{df}{dt}$= 0    &&&&&&&&&&\\    
          & $\frac{d^2f}{dt^2}$=0 &&&&&&&&&&\\  
\hline
\hline
LINE A    & $f_{0}$=\unit[50]{Hz}             & 250 & 50-200  &  1.12  & 1.25 &     0.64 & 3.01 & 2.45e-08 & 3.08e-08 & 8.05e-09 & 1.78e-07\\ 
          & $\frac{df}{dt}$=\unit[0.6]{Hz $\mathrm{s^{-1}}$}    &&&&&&&&&&\\   
          & $\frac{d^2f}{dt^2}$=0 &&&&&&&&&&\\    
\hline
LINE B    & $f_{0}$=\unit[900]{Hz}            & 100 &   700-900 &     1.62  & 1.28 &     0.76 & 1.60 & 1.67e-06 & 1.04e-06 & 3.62e-07 & 1.63e-06\\ 
          & $\frac{df}{dt}$=\unit[-2]{Hz $\mathrm{s^{-1}}$}    &&&&&&&&&&\\
          & $\frac{d^2f}{dt^2}$=0  &&&&&&&&&&\\
\hline
\hline
QUAD A    & $f_{0}$=\unit[50]{Hz}              &  30 &    50-200 &     0.83  & 0.75 &    0.66 & 1.81 & 9.02e-09 & 7.34e-09 & 5.72e-09 & 4.28e-08 \\ 
          & $\frac{df}{dt}$=0      &&&&&&&&&&\\
          & $\frac{d^2 f}{dt^2}$=\unit[0.33]{Hz $\mathrm{s^{-2}}$} &&&&&&&&&&\\
\hline
QUAD B    & $f_{0}$=\unit[500]{Hz}                  &  70 &   500-600 &     1.21  & 1.07 &     0.75 & .96 & 4.43e-07 & 3.48e-07 & 1.69e-07 & 2.76e-07\\ 
          & $\frac{df}{dt}$=0           &&&&&&&&&&\\
          & $\frac{d^2 f}{dt^2}$=\unit[0.04]{Hz $\mathrm{s^{-2}}$} &&&&&&&&&&\\

\hline
\hline
SG A      & $f_{0}$=\unit[90]{Hz}           & 150 &        90 & -         & 5.50 & -        & 3.42 & -        & 2.84e-07 & -        & 1.10e-07\\ 
          & $\tau$=\unit[30]{s} &&&&&&&&&&\\
\hline
SG C      & $f_{0}$=\unit[405]{Hz}          & 250 &       405 & -         & 3.79 & -        & 1.95 & -        & 2.57e-06 & -        & 6.81e-07\\ 
          & $\tau$=50\unit[50]{s} &&&&&&&&&&\\

\hline
\hline
WNB A     & -                &  20 &    50-400 &     2.86  & 2.04 & -        & 4.74 & 5.17e-07 & 2.63e-07 & -        & 1.42e-06\\ 
\hline
WNB B     & -                &  60 &   300-350 &     2.93  & 1.97 & -        & 1.73 & 1.80e-06 & 4.52e-07 & -        & 3.49e-07\\ 
\hline
WNB C     & -                & 100 &   700-750 &     5.36  & 3.20 & -        & -        & 1.53e-05 & 5.45e-06 & -        & - \\ 
\hline
\hline
\end{tabular}
\caption{Search sensitivity of the four pipelines to the 22 waveform families used to cover the unmodeled long transient parameter space. The $h_{rss}$ at 50\% efficiency is computed for a FAR of 1 event in 50 years. $E_{GW}^{50\%}$ is the GW energy emitted by a source located at 10 kpc for which the search efficiency drops below 50\% for a FAR of 1 event in 50 years. The models are not sequentially named to avoid confusion with models used in former studies. The second column provides the parameters of the waveforms as defined in section~\ref{sec:waveforms} or in~\cite{Abbott:2015vir}.}
\label{table:LT_cmp}
\end{table*}

%% file: discussion.tex
\section{Search results}\label{sec:discussion}

\begin{figure*}[hbtp!]
  \centering
  \includegraphics[height=5in]{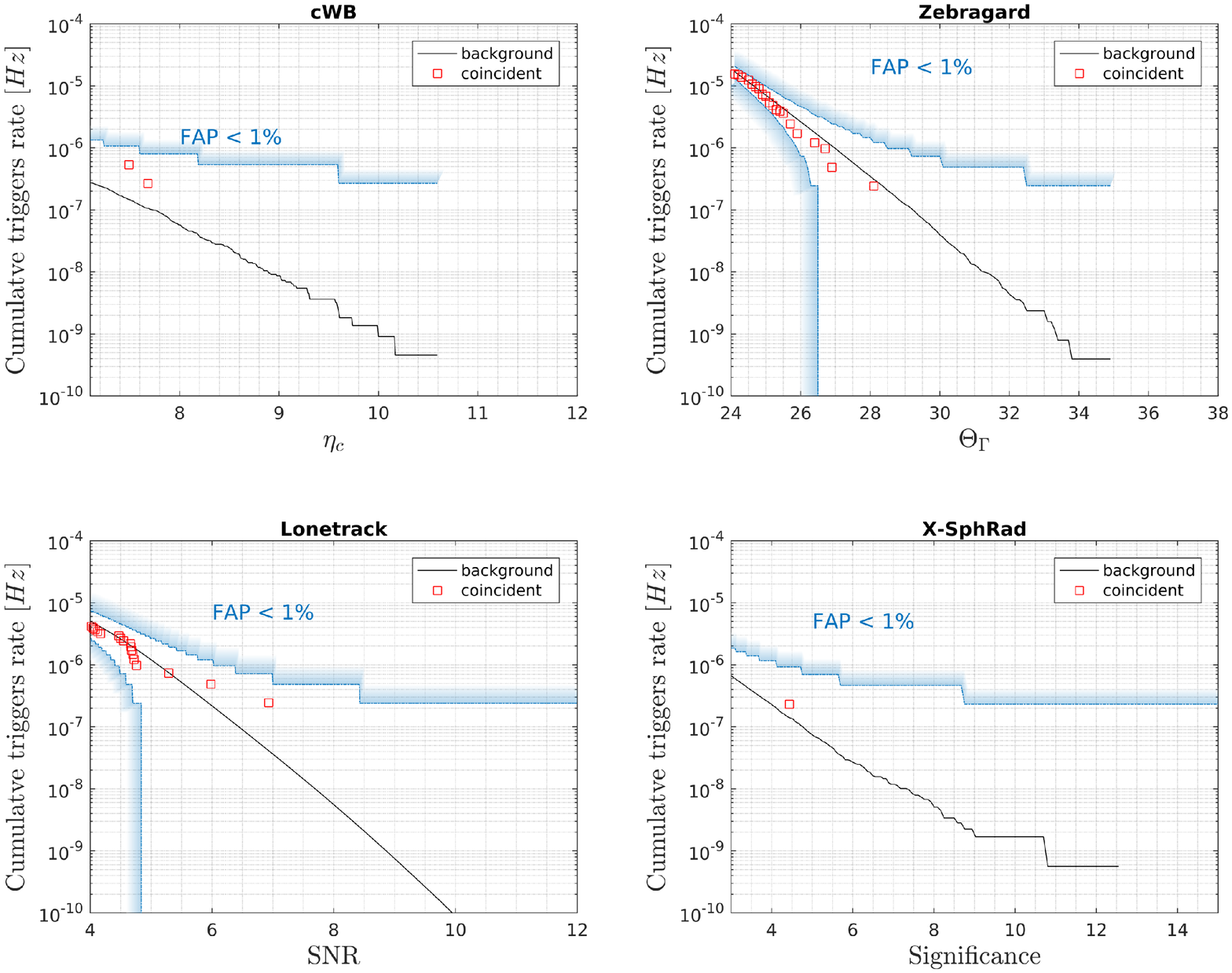}
  \caption{Cumulative trigger rate as function of the triggers' ranking statistic for the four pipelines. The coincident triggers are represented by the red squares while the black curves show the estimation of the contribution of the accidental coincident noise triggers. The blue isocurves indicate the trigger rate that corresponds to a false alarm probability (FAP) lower than 1\%. This illustrates that all coincident triggers' distributions are compatible with the expected background. For cWB and X-SphRad the lower isocurve is not displayed because it falls outside of the axis limits.}
  \label{fig:FAR}
\end{figure*}

Figure~\ref{fig:FAR} shows the distributions of the cumulative rate of coincident data triggers for each pipeline; these are ranked according to the pipelines' detection statistic and are shown together with the estimated background. The X-SphRad and cWB distributions contain fewer triggers than the Zebragard or Lonetrack distributions because of the selection criteria that remove many low significant triggers at early stages. No significant excess of coincident triggers is found by any pipeline. The properties of the most significant triggers from each pipeline are reported in Table~\ref{tab:LT_events}. They are all compatible with the O1 background expectations as underlined by the rather large values of their false alarm probabilities (FAP). The FAP is the probability of observing at least one background trigger with a ranking statistic larger than a given threshold.

\input{tabevents}

Given the absence of long-duration transient GWs in the O1 data, we have updated the limits established in~\cite{Abbott:2015vir}. Assuming a Poissonian distribution of long-duration GW sources, we compute the 90\% confidence level limit of the trigger rate using the loudest event statistic method~\cite{Brady:2004gt}, where systematic uncertainty coming from the strain amplitude calibration is folded into the upper limit calculation as in~\cite{Abbott:2015vir}. During the O1 science run, the amplitude calibration uncertainty was measured to be 6\% and 17\% for the H1 and L1 detectors, respectively, in the \unit[24--2048]{Hz} frequency band~\cite{Abbott:2016jsd}.

\begin{figure}[t]
  \centering
  \includegraphics[height=2.5in]{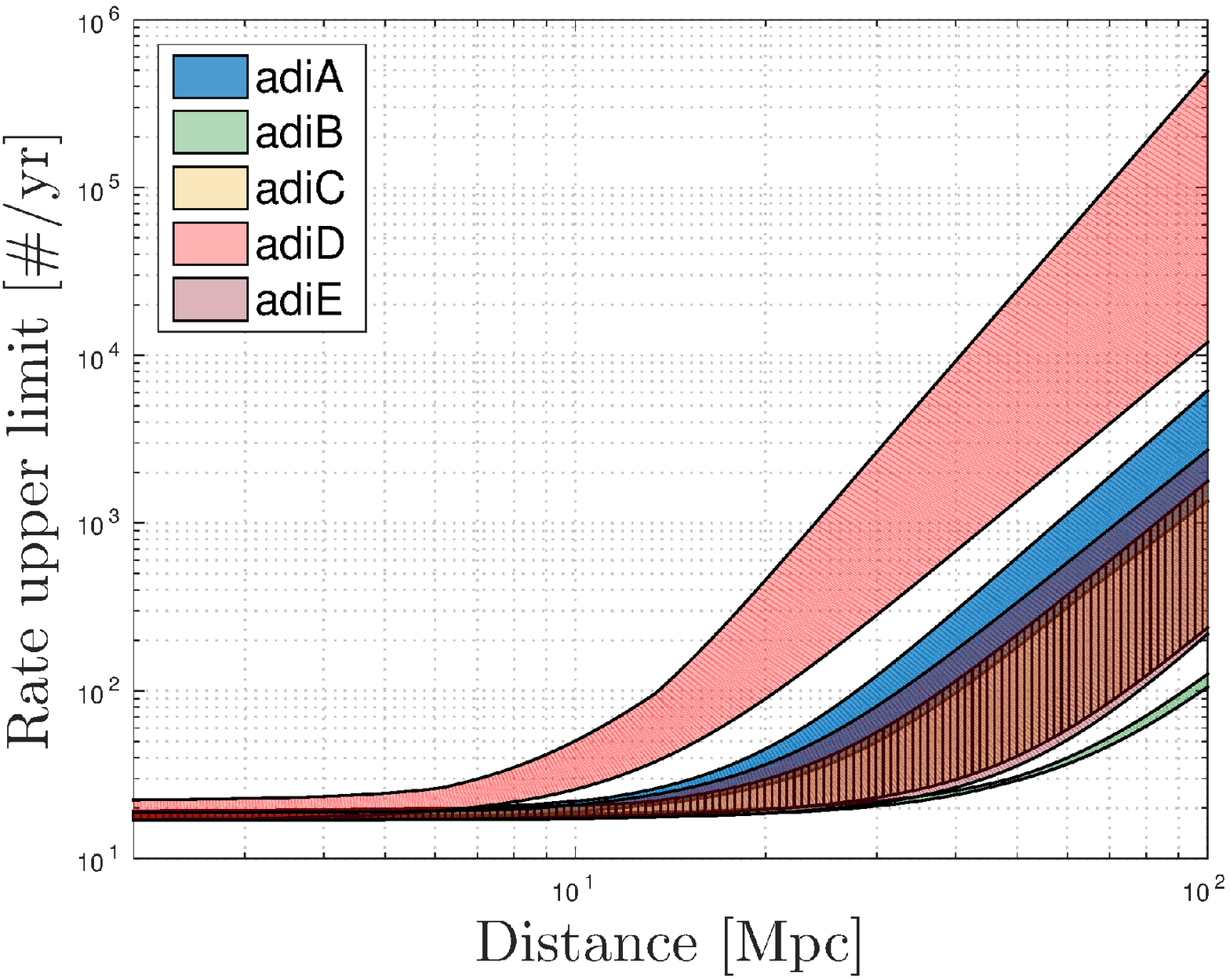}
  \caption{Upper limits at 90\% confidence on the rate of GW events from accretion disk instability as a function of the distance. The band covers the results from the best and the worst pipelines for each tested waveforms. O1 amplitude calibration errors are accounted for in the upper limits calculation.}
  \label{fig:rateGW}
\end{figure}

Figure~\ref{fig:rateGW} shows the rate upper limit as a function of distance for the ADI signals. The area is defined by the most and the least performing pipelines. The exclusion rate at short distance is limited by the observational duration. Since O1 is shorter than S5 or S6, the event rate is less constrained. Conversely, the maximal distance for which one can expect to detect an event is improved by a factor $\sim$ 3.

Distances at which we can detect a signal with 50\% efficiency are compared for all astrophysical waveforms in Table~\ref{table:LT_dist}. As already seen, detection distances for the 5 ADI waveforms are between \unit[10 -- 60]{Mpc}. On the other hand, the chance of detecting GWs from a magnetar or from the accretion of a black hole is limited to sources in the Local Group.

\input{tabdist}

The fact that we do not see any signals in O1 is not unexpected. First, O1 is a short run, with only \OOneLivetime~ of total coincident data, which is enough to detect multiple coalescences of binary black holes but quite short to detect long-duration GW signals considering the large uncertainties or unknowns about the rates of each of the potential long transient GW sources. Next, the energy of a long-duration signal is spread over a large number of pixels, which causes a decrease in the sensitivity of the pipelines. This explains why the short transient O1 search~\cite{Abbott:2016ezn} is roughly an order of magnitude more sensitive at a given frequency. Nevertheless, when compared to the S5/S6 results~\cite{Abbott:2015vir}, the O1 long-duration transient search is better by a factor $\sim$ 10 due to the improvements in detector sensitivities.

%% file: tabevents.tex
\renewcommand{\arraystretch}{1}
\begin{table}[h]
\begin{tabular}{ccccc}
\hline
\hline
Pipeline & Ranking   & FAP & Frequency & Duration\\
GPS time & statistic &           & [Hz]      & [s]\\
\hline
\hline
cWB & $\eta_{c}=7.6$                    & 0.33 & 2039-2041 & 5.5\\
1132990790 & & &\\
\hline
Zebragard & $\Theta_{\Gamma}=28.2$          & 0.72 & 1034-1120 & 51\\
1131758576 & & &\\
\hline
Lonetrack & $\mathrm{SNR=6.95}$                  & 0.36 & 85-1549  & 208\\
1136368706 & & &\\
\hline
X-SphRad & $\mathrm{Significance}=4.5$ & 0.44 & 895-909   & 4\\
1135861536 & & &\\
\hline
\hline
\end{tabular}
\caption{Properties of the most significant coincident triggers found by each of the long transient search pipelines during the O1 observational run. FAP is the probability of observing at least 1 noise trigger more significant that the most significant coincident trigger.}
\label{tab:LT_events}
\end{table}

%% file: tabdist.tex
\begin{table}
\begin{tabular}{ccccc}
  \hline
  \hline
\multirow{3}{*}{Waveform}    & \multicolumn{4}{c}{$\mathrm{Distance}^{50\%}$ [Mpc]}\\
\cline{2-5}                  & cWB & \multicolumn{2}{c}{STAMP-AS} & X-SphRad \\ 
                             &  & Zebragard     & Lonetrack    & \\
\hline
\hline
FA A       & 1.08 & 1.34 & 1.69 & -\\
\hline
FA B       & 1.76 & 1.91 & 3.32 & -\\
\hline
\hline
ADI A      & 19.1 & 17.1 & 22.0 & 23.6\\ 
\hline
ADI B      & 58.3 & 54.6 & 52.5 & 54.4\\ 
\hline
ADI C      & 29.1 & 30.5 & 41.1 & 22.6\\ 
\hline
ADI D      & 10.1 & 8.33 & 12.3 & 8.02\\ 
\hline
ADI E      & 33.6 & 39.2 & 46.0 & 19.1\\ 
\hline
\hline
magnetar D & 0.14 & 0.11 & 0.19 & -\\
\hline
magnetar E & 0.20 & 0.20 & 0.37 & - \\
\hline
magnetar F & 0.50 & 0.57 & 1.02 & 0.68\\
\hline
magnetar G & 0.43 & 0.35 & 0.61 & 0.71\\
\hline
\hline
\end{tabular}
\caption{Distances at which the pipeline efficiency drops below 50\% for a FAR of 1 event in 50 years for the accretion disk instability, magnetar and fallback accretion signals considered in the O1 search.}
\label{table:LT_dist}
\end{table}

%% file: conclusion.tex
\section{Conclusion}\label{sec:conclusion}

This paper reports the results of an all-sky search for unmodeled long-duration transient GWs in the first Advanced LIGO observing run. The parameter space covered by this search has been increased compared to the preceding search. Four different pipelines have searched for GW signals to efficiently cover the large space of possible waveforms. The most significant triggers found by each pipeline are consistent with the noise background, excluding for now a long duration GW transient detection. \\

Upper limits have been set on the rate of events for three families of long-duration GW transients (fallback accretion on neutron stars, black hole accretion disk instabilities and magnetar giant flares). They indicate we are sensitive to potential sources in the Local Group. Alternatively, if we consider a source in the Galaxy (\unit[10]{kpc}) we are sensitive to sources emitting at least \unit[$6\times10^{-9}$]{\msuncd} for frequencies where the detectors' sensitivities are maximal. This is a lower bound (our results are spread over almost two orders of magnitude) but this is still an interesting achievement as it addresses an energy range that is astrophysically relevant \cite{Turolla:2015mwa,Mueller:2003fs}. New data have been acquired recently by the LIGO detectors (observing run O2) with a sensitivity similar to O1 and a longer observation time which increases the chance of observing a long-duration transient GW source~\cite{Aasi:2013wya}. The Advanced Virgo detector has joined for the first time the advanced GW detector network on August $\mathrm{1^{st}}$ 2017; this increases sky coverage and improves the prospects for detection. In a few years, Advanced LIGO and Advanced Virgo should reach their design sensitivities. We have shown that we should gain between one and two orders of magnitude, depending on the frequency range, in the sensitivity to detect GW energy as low as $\sim$ $10^{-8}$ \msuncd\, for a source emitting a monochromatic signal at $\sim$ \unit[90]{Hz} and located at \unit[10][kpc].

%% file: LVCacknowledgments.tex
The authors gratefully acknowledge the support of the United States
National Science Foundation (NSF) for the construction and operation of the
LIGO Laboratory and Advanced LIGO as well as the Science and Technology Facilities Council (STFC) of the
United Kingdom, the Max-Planck-Society (MPS), and the State of
Niedersachsen/Germany for support of the construction of Advanced LIGO
and construction and operation of the GEO600 detector.
Additional support for Advanced LIGO was provided by the Australian Research Council.
The authors gratefully acknowledge the Italian Istituto Nazionale di Fisica Nucleare (INFN),
the French Centre National de la Recherche Scientifique (CNRS) and
the Foundation for Fundamental Research on Matter supported by the Netherlands Organisation for Scientific Research,
for the construction and operation of the Virgo detector
and the creation and support  of the EGO consortium.
The authors also gratefully acknowledge research support from these agencies as well as by
the Council of Scientific and Industrial Research of India,
the Department of Science and Technology, India,
the Science \& Engineering Research Board (SERB), India,
the Ministry of Human Resource Development, India,
the Spanish  Agencia Estatal de Investigaci\'on,
the Vicepresid\`encia i Conselleria d'Innovaci\'o, Recerca i Turisme and the Conselleria d'Educaci\'o i Universitat del Govern de les Illes Balears,
the Conselleria d'Educaci\'o, Investigaci\'o, Cultura i Esport de la Generalitat Valenciana,
the National Science Centre of Poland,
the Swiss National Science Foundation (SNSF),
the Russian Foundation for Basic Research,
the Russian Science Foundation,
the European Commission,
the European Regional Development Funds (ERDF),
the Royal Society,
the Scottish Funding Council,
the Scottish Universities Physics Alliance,
the Hungarian Scientific Research Fund (OTKA),
the Lyon Institute of Origins (LIO),
the Paris \^{I}le-de-France Region,
the National Research, Development and Innovation Office Hungary (NKFI),
the National Research Foundation of Korea,
Industry Canada and the Province of Ontario through the Ministry of Economic Development and Innovation,
the Natural Science and Engineering Research Council Canada,
the Canadian Institute for Advanced Research,
the Brazilian Ministry of Science, Technology, Innovations, and Communications,
the International Center for Theoretical Physics South American Institute for Fundamental Research (ICTP-SAIFR),
the Research Grants Council of Hong Kong,
the National Natural Science Foundation of China (NSFC),
the Leverhulme Trust,
the Research Corporation,
the Ministry of Science and Technology (MOST), Taiwan
and
the Kavli Foundation.
The authors gratefully acknowledge the support of the NSF, STFC, MPS, INFN, CNRS and the
State of Niedersachsen/Germany for provision of computational resources.